# Impact of the *fac/mer* isomerism on the excited states dynamics of pyridyl-carbene Fe(II) complexes

*Kevin Magra,[a] Edoardo Domenichini,[b] Antonio Francés-Monerris,[c] Cristina Cebrian,[a] Marc Beley,[d] Mohamed Darari,[d] Mariachiara Pastore,[c] Antonio Monari,[c] Xavier Assfeld,[c] Stefan Haacke[b] and Philippe C. Gros\*,[d]*

a) Université de Lorraine, CNRS, L2CM, F-57000 Metz, France. b) Université de Strasbourg, CNRS, IPCMS, F-67000 Strasbourg, France. c) Université de Lorraine, CNRS, LPCT, F-54000 Nancy, France, d) Université de Lorraine, CNRS, L2CM, F-54000 Nancy, France

AUTHOR INFORMATION

**Corresponding Author**

*Philippe C. Gros: philippe.gros@univ-lorraine.fr

ABSTRACT

The control of photophysical properties of iron complexes and especially of their excited states decay is a great challenge in the search for sustainable alternatives to noble metals in photochemical applications. Herein we report the synthesis and investigations of the photophysics of mer and fac iron complexes bearing bidentate pyridyl-NHC ligands, coordinating the Fe with three ligand-field enhancing carbene bonds. Ultrafast transient absorption spectroscopy reveals two distinct excited state populations for both *mer* and *fac* forms, ascribed to the populations of the T1 and the T2 states, respectively, which decay to the ground state via parallel pathways. We find 3-4 ps and 15-20 ps excited state lifetimes, with respective amplitudes depending on the isomer. The longer lifetime exceeds the one reported for iron complexes with tridentate ligands analogues involving four iron-carbene bonds. By combining experimental and computational results, a mechanism based on the differential trapping of the triplet states in spin-crossover regions is proposed for the first time to explain the impact of the fac/mer isomerism on the overall excited-state lifetimes. Our results clearly highlight the impact of bidentate Pyridyl-NHC ligands on the photophysics of iron complexes, especially the paramount role of fac/mer isomerism in modulating the overall decay process, which can be potentially exploited in the design of new Fe(II)-based photoactive compound.

INTRODUCTION

The replacement of platinoids (Pt, Ru, Ir) by other more accessible metals in photoactive systems constitutes a great scientific and technological challenge. This accomplishment is crucial for the development of cheaper and more eco-friendly light-driven devices such as organic light-emitting diodes (OLEDs)[1] or dye-sensitized solar cells (DSSCs).[2] As a matter of fact, in the DSSCs field, ruthenium complexes have dominated the scene and boosted the technological development thanks to their optimal electrochemical and photophysical properties, such as an efficient coverage of the solar spectrum and a fast electron transfer via long-lived metal-to-ligand charge transfer (MLCT) excited states.[3,4] However, the low abundance, high price and non-negligible toxicity of ruthenium definitely represent a major limit in view of large-scale technological applications. Thus, the quest for sustainable alternatives has pushed the scientific community to develop analogous iron-based compounds, the latter being a cheaper, more abundant and less toxic element, as it was pioneered by Ferrere two decades ago.[5,6]

Nevertheless, the population of low-energy metal-centered (MC) states in typical iron polyimine complexes leads to the ultra-fast deactivation of the MLCT states initially accessed upon light excitation of the complex and as a consequence to the loss of the suitable photophysical properties.[7,8]

Hence, the development of iron-based optically active devices will necessarily imply the synthesis and production of systems presenting long-lived MLCT states. In turn, this task is a veritable scientific challenge and necessitates to achieve a thorough understanding of the fine photophysical process dictating the MLCT deactivation and most importantly the non-trivial coupling between electronic and vibrational effects. In this respect, the fine comprehension of the fundamental factors tuning the excited state relaxation and of their subtle interplay is fundamental.

Considerable success in increasing the excited-state lifetime of organometallic iron complexes has been achieved in the very last years thanks to the use of *N*-heterocyclic carbenes (NHC) as ligands. Indeed, the combined strong σ-donating and weak-to-moderate π-accepting character of NHC induces a higher splitting of the iron d orbitals, resulting in a destabilisation of the MC states over the MLCT manifold and consequently a hampering of the deactivation pathways. For instance, the use of tridentate pyridyldicarbenes ligands allowed achieving a remarkable $^3$MLCT lifetime of 9 ps for complex **C0** (Fig. 1).[9] Confirming the subtle interplay between different structural and chemical effects, our group also showed[10] that the presence of a carboxylic group on the central pyridyl ring is non-innocent and almost doubled the $^3$MLCT lifetime up to 16.5 ps.[11] Switching from imidazolylidenes to benzimidazolylidenes moieties in the chromophoric ligand further increased the excited state lifetime, attaining a record lifetime of 26 ps.[12] This experimental observation has also been rationalized by molecular modelling revealing, for the first time, an inversion of the MLCT/MC energy levels for an iron complex. In addition, the inclusion of carboxylic group allowed the sensitization of $TiO_2$ surface and the production of the first working DSSC device based on iron NHC, although with a low efficiency of light-to-electricity conversion.[11,13]

An alternative strategy to increase the ligand-field splitting, and hence the MLCT lifetime, is to employ ligands that reduce the angular strain around the metal centre upon coordination.[14,15,16,17,18,19] This effect is invoked to justify the different photophysical behaviour of the angularly strained $[Ru(tpy)_2]^{2+}$ complex (tpy = 2,2':6',2''-terpyridine) that experiences an ultra-fast non-radiative via MC states population,[20] and the weakly distorted bidentate homologue $[Ru(bpy)_3]^{2+}$ (bpy = 2,2'-bipyridine)[21] that is instead strongly luminescent at room temperature since the deactivation channels are not accessible.[22] As a consequence, the selection of tridentate or bidentate coordinating ligands can have a deep impact on the photophysical properties of the final complex.

However, the photophysical phenomena taking place in transition metal complexes are extremely complicated also due to the high density of excited states and their high coupling. Therefore, many different relaxation channels may compete, and a simple rationalization based on very intuitive concepts, such as the ligand-field splitting, although useful, does not allow to catch all the subtle details of the processes. More elaborate modelling strategies, involving the full exploration of the potential energy surfaces (PESs) of the involved states are needed to complete the picture.

Recently, and to influence the angular strain, we have reported the first example of iron complex with bidentate pyridylcarbene ligands **C1**.[23] The asymmetric nature of the employed ligand led to different geometrical isomers. In fact, **C1** was obtained as a mixture of mostly the meridional (*mer*) isomer with a concomitant fraction of the facial (*fac*) isomer in a 14:1 ratio. A thorough molecular modelling of the different singlet, triplet and quintet excited states PES involved in the photoresponse evidenced totally different landscapes for the two isomers. In particular, the low-lying triplet PES of the *mer* isomer, leading to the MC population and deactivation, was found to be much steeper than the one of the corresponding *fac* isomer.[23] This aspect is intimately linked to the larger stabilization of the $^3$MC state in the *mer* arrangement as compared to the *fac* one. Altogether, our molecular modelling results indicate a much faster deactivation for the *mer* isomer.

This result represented, to the best of our knowledge, the first hypothesis on the possibility of tuning and controlling the excited states life-time by the effects of fac/mer isomerism. As such, it also represents a fundamental aspect of the inherent photophysics of organometallic compounds that should be precisely investigated and whose implication can go far beyond the production of photoactive iron complexes.

However, the two isomers could not be separated, hence precluding the possibility to perform time-resolved spectroscopy to obtain the two separate life-times. To overcome this problem, we propose a different strategy based on the precise control of the geometrical arrangement of the ligands to selectively obtain only one of the two isomers. This leads to the synthesis of a new *C$_3$*-symmetric hemicage ligand **L2** based on bidentate pyridine-imidazolylidene moieties, subsequently used to prepare the corresponding pure *fac* iron(II) complex **C2** (Fig. 1).

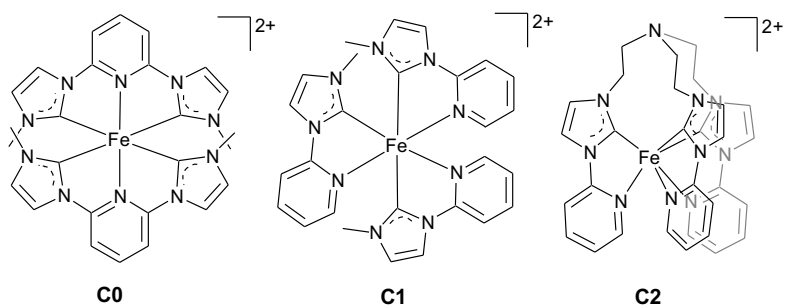

**Figure 1.** Complexes studied in this work. In all cases, counter ion is $PF_6^-$.

In this contribution we report the full characterization of both **C1** and **C2** including molecular modelling and time-resolved spectroscopy, thereby allowing for the quantification and the rationalisation of the *fac/mer* isomerism on the properties of these iron-based compounds. Most importantly, while globally confirming the difference in relaxation times of the two isomers, we also show how this effect is due to a quite complex interplay between two decay pathways that are, in turn, affected differentially by the PES topologies.

This work represents the first direct observation of different decay characteristic times in *fac* and *mer* isomers of iron compounds. As such, it opens the way to a novel strategy that can be synthetically pursued to achieve the precise tuning of important photophysical properties of materials.

RESULTS AND DISCUSSION

**Synthesis and characterization**

The preparation of pyridylimidazolium salts precursors **L1** and **L2** was achieved by reacting 2-(1*H*-imidazol-1-yl)pyridine **1** with the appropriate haloalkane. In the case of **L1**, the quarternisation reaction was performed with methyl iodide in acetonitrile at 85ºC (Scheme 1).[24,25] As for **L2**, tris(2-chloroethyl)amine **2** was selected as the alkylating agent. Temperatures as high as 150 °C and isopropanol as solvent were required to attain the triple-fold substitution. It is noteworthy to mention the dramatic effect of the microwave heating in both substitution reactions, which were performed in only 1h while similar reactions performed under classical thermal conditions have been previously reported to last up to 7 days.[26]

The target complexes **C1** and **C2** were obtained following a recently reported protocol by our group. Upon *in situ* generation of the carbene species by deprotonation of the pyridylimidazolium salts in the presence of *t*-BuOK, the resulting ligand is then coordinated to iron *via* $FeCl_2$ metal precursor in DMF (Scheme 1). It should be noted the importance of adding the base at the end in order to reduce possible detrimental side-reactions, *e.g.* carbene degradation or iron precipitation.[26,27] In the case of tripodal ligand **L2**, the protocol required more diluted conditions in order to reduce the formation of oligomeric side-products. Successful coordination was confirmed in both cases by analysing the variation of the chemical shifts in $^1$H-NMR spectra. As expected, the pro-carbenic proton was absent in the final complexes. Moreover, the coordination of the three pyridine rings in **C2** was nicely confirmed by the noticeable shift change of the alpha pyridine hydrogens, excluding a possible coordination of the central amine moiety (See Fig. S1).

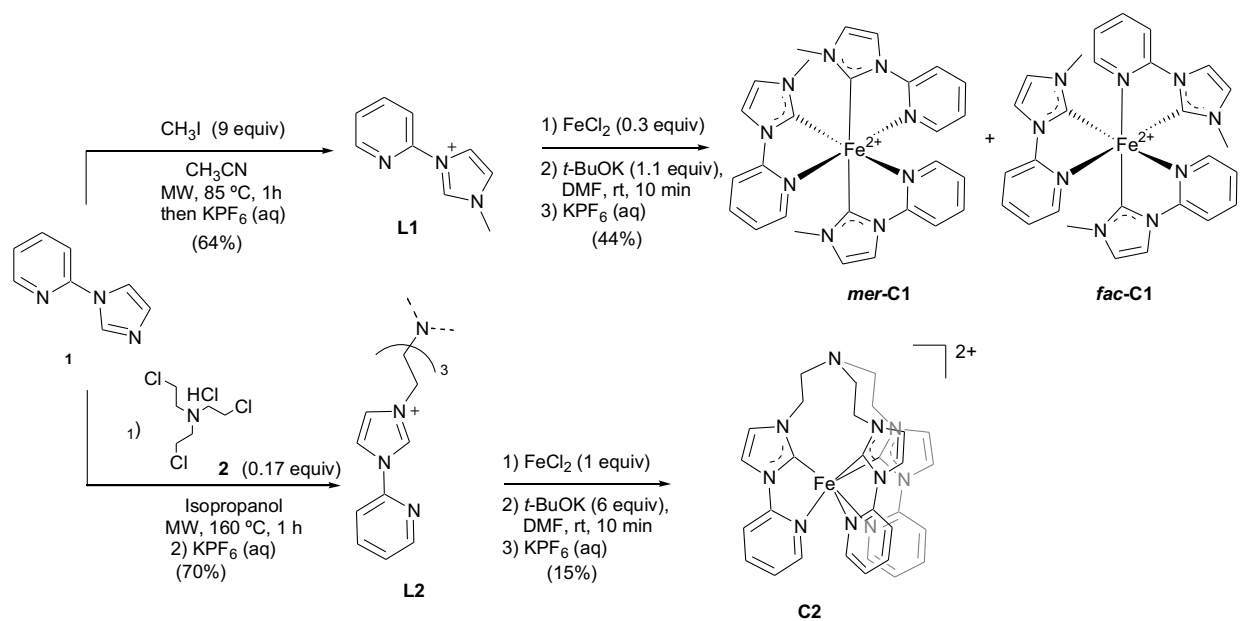

**Scheme 1.** Overall synthesis of pro-carbene ligands **L1** and **L2**, together with final iron(II) complexes **C1** and **C2**. Counterion is $PF_6$ for all charged compounds.

As aforementioned, **C1** was obtained as a non-separable mixture of *mer* and *fac* isomers in a 14:1 ratio. This remarkably high ratio might be attributed to the steric requirements of the ligand, that would particularly disfavour a facial arrangement, along with the employed reaction conditions.[28,29] Nevertheless, the identification and characterisation of both isomers was carried out unambiguously by NMR analysis due to their differences in signal intensity and pattern. Concerning **C2**, the three bidentate units are magnetically equivalent, which correlates well with the inherent $C_3$ symmetry of the complex. It is worth noting that the constrained geometry causes the four hydrogens of the ethylene units to be non-equivalent. Despite all our attempts, we were unsuccessful in growing quality crystals suitable for diffraction XRay analysis so we have resorted to high-level molecular modelling to get the molecular structures of **C0**, **C1** and **C2.** The ground state equilibrium geometry of the tridentate and bidentate compounds has been obtained by molecular modelling (Fig. 2) and the details are reported (Table S1).

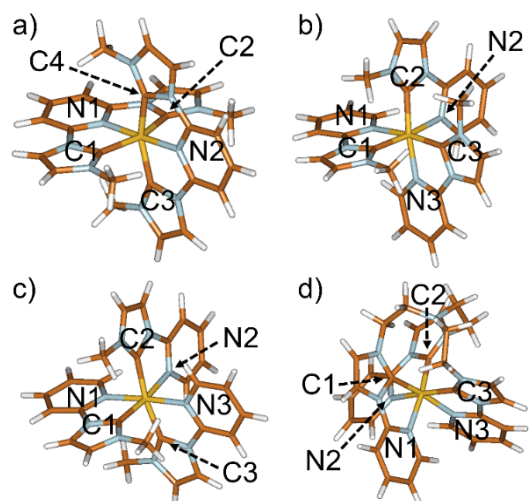

**Figure 2.** Atom labelling and ground-state equilibrium geometries for a) **C0**, b) *fac*-**C1**, c) *mer*-**C1** and d) **C2**. The optimization procedures have been conducted with very tight convergence thresholds

**C0** with the tridentate ligand has shorter distances between the iron atom and the coordinating nitrogen atoms and slightly longer Fe-C distance. On the other hand, the impact on the angular strain measured via the N-Fe-C bite angles appears marginal, ~~C0 having smaller bite angles (less than 2°)~~. Most importantly the geometry of **C2** is globally equivalent to the one of the *fac*-**C1** isomer.

**Electronic and electrochemical properties.**

The redox properties of both complexes were analyzed using cyclic voltammetry with SCE as standard electrode (Table 1 and Fig.Sx). At positive potentials, **C1** and **C2** displayed a one-electron reversible Fe(II) to Fe(III) oxidation wave at 0.65 V and 0.61 V, respectively. The electronic transfer is slightly easier than for **C0** ($E_{1/2}$= 0.71 V).[9] Thus, the HOMO energy level is slightly higher for **C2** than for **C1** and **C0**, but much higher than those of $[Fe(bpy)_3]^{2+}$ ($E_{1/2}$= 1.05 V)[30] and $[Fe(tpy)_2]^{2+}$ ($E_{1/2}$= 1.13 V).[31] This behaviour is due to the strong σ-electron-donating character of the carbene-metal bond. In the negative potential domain, irreversible reduction processes occurred at about -2.00 V/SCE for the three complexes. This process corresponds to the injection of electrons in the antibonding π* orbital corresponding, to a first approximation, to the LUMO orbital of the complex. For **C0** the reduction process is mono-electronic, ($i_{pc}/i_{pa}$ = 1) where $i_{pa}$ is associated to the oxidation transfer in the positive potential domain and $i_{pc}$ corresponds to the irreversible transfer at -2.00 V. For **C1** and **C2**, two mono-electronic reduction waves are distinguished. The reduction potentials of irreversible processes are shifted by about 0.8V to lower values by comparison with [Fe(bpy)$_3$]$^{2+}$ ($E_{1/2}$= -1.30V)[30] and [Fe(tpy)$_2$]$^{2+}$ ($E_{1/2}$=-1.20 V).[31] This important shift can be explained by a combined σ-donor character of the ligand and the π-back donation of Fe(II).

The complexes were also characterized by UV-vis spectroscopy (Fig. 3 and Table 1). For all the systems, the absorption spectrum, shows three bands in the 200-700 nm region. Usually, the intense bands near 250-300 nm can be assigned to intraligand (IL) [π→π*] transitions. In the lower energy region, the two less intense, broader bands are attributed to MLCT transitions as previously reported for **C0**,[9] **C1**,[23] and other related Fe(II) complexes. The band in the 450-500 nm area is clearly blue shifted by 30 nm for **C1** and **C2** compared with **C0**.

**Table 1.** Electronic and electrochemical properties of complexes

| Complex | $E_{½\,ox}$ (Fe$^{III}$/Fe$^{II}$) (V) [b] [$E_{pa}$-$E_{pc}$] (mV) | $E_{red}$ (V) [b] | $\Delta E$ (eV) [c] | $\lambda_{abs-max}$ (nm) [$\varepsilon$(M$^{-1}$.cm$^{-1}$)] [a] |
|---|---|---|---|---|
| **C0** | 0.71 (rev) [90] | -2.00 (irrev) | 2.71 | 287 [31400] 393 [9000] 460 [15900] |
| **C1** | 0.65 (rev) [90] | -1.97 (irrev) -1.94 (irrev) | 2.62 2.59 | 272 [26000] 360 [4500] 430 [12000] |
| **C2** | 0.61 (rev) [90] | -1.94 (irrev) -2.04 (irrev) | 2.55 2.65 | 273 [19500] 369 [5200] 438 [8000] |

[a] Measured in CH$_3$CN at 25 °C. [b] First oxidation potential. Potentials are quoted *vs* SCE. Under these conditions, $E_{1/2\,(Fc+/Fc)}$ = 0.38V/S.C.E. Recorded in CH$_3$CN using Bu$_4$N$^+$PF$_6^-$ (0.1M) as supporting electrolyte at 100 mV. s$^{-1}$. [c] Electrochemical band gap ($\Delta E = E_{½\,ox} - E_{pred1}$).

The optical properties have also been modelled by TD-DFT allowing also to characterize the nature of the electronic excited states. Figure S13 displays the calculated UV-Vis spectrum of **C2**, which shows an intense band centred at λ ~443 nm, in reasonably good agreement with the experimental recordings (see Fig. 3). Note that the small red shift of **C2** with respect to **C1** observed in the experimental band maxima (see Fig. 3) is also captured by our theoretical protocol, since the convoluted spectrum of the 1:14 *fac/mer*-**C1** mixture has an absorption band peaking at ~415 nm.[23]

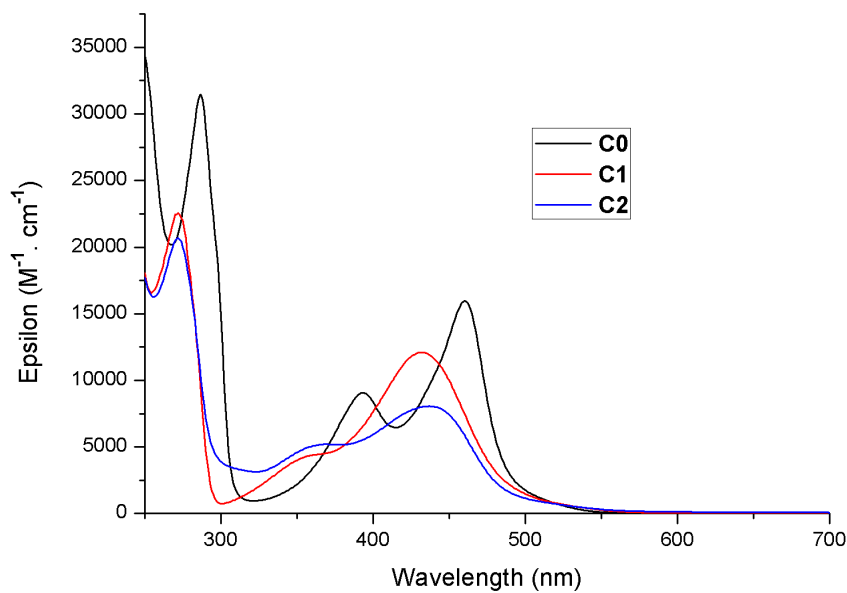

**Figure 3.** UV-Vis spectra of **C0**, **C1** and **C2** in in air-equilibrated CH$_3$CN solution

This is indeed confirmed by the analysis of the excited states topology (Fig. S14). Interestingly, **C2** has additional electronic transitions from the nitrogen atom of the 'bridge' amino group to the π* orbitals of the pyridyl-carbene ligands (n$_N$L). The excitation energies lie in the 350 – 450 nm range, although the associated oscillator strengths are small as compared to the MLCT transitions (see Fig. S15). However, certain MLCT/n$_N$L mixing, also enhanced by the symmetry breaking induced by the tertiary amine 'bridge', might be responsible for the absorption intensity loss displayed by **C2** in that region, with respect to the more efficient absorption of **C1.**

**Excited-state computational studies**

As already stated, important differences in the excited states PES landscapes have been found between *fac* and *mer* isomers.[23] Hence, it is necessary to prove that the **C2** complex represents a good model for a pure *fac*-FeNHC complex and does not alter this energetic landscape.

The PESs and the associated geometrical distortions that drive the **C2** decay to the ground state have been modelled using DFT and TD-DFT methods. Given the ultrafast population of the triplet manifold determined for this compounds and other Fe(II) derivatives in the literature,[8,18,32,33,34,35] it is thus expected that the triplet states can be accessed already at the Franck-Condon or $S_1$ ($^1$MLCT) equilibrium geometry, taking also into account the structural similarity of both geometries (see Table S2), as it is also analysed in our previous report.[23] The dark $n_N$,L states shown in Figure S15 will likely be populated via the cascades of internal conversions and intersystem crossings, even though minor relevance of these states on the excited-state decay mechanisms is predicted since the lowest-energy singlet and triplet states are of MLCT/MC nature, as detailed in the following.

Fig. 4 tracks the low-lying singlet, triplet and quintet states along the key structures involved in the decay process, namely the $S_0$ min, $S_1$ ($^1$MLCT) min, $T_1$ ($^3$MC) min and $Q_1$ min geometries. As in the case of *fac*-**C1**,[23] the triplet PESs of **C2** evolve adiabatically from MLCT to MC nature as the system approaches to the $T_1$ ($^3$MC) min structure. The natural transition orbitals (NTOs) isosurfaces for the lowest-lying singlet and triplet states are shown in Figure S16. Coherently with what observed for *fac*-**C1** and *mer*-**C1**,[23] the main coordinate driving the process is the enlargement of the Fe-N1 bond, which has a bond distance of 2.82 Å at the $T_1$ ($^3$MC) min geometry (Fig. 4a). Further Fe-N1 stretching is required to reach the $T_1$-$S_0$ singlet-triplet crossing (STC) at 3.05 Å (Fig. 4b).

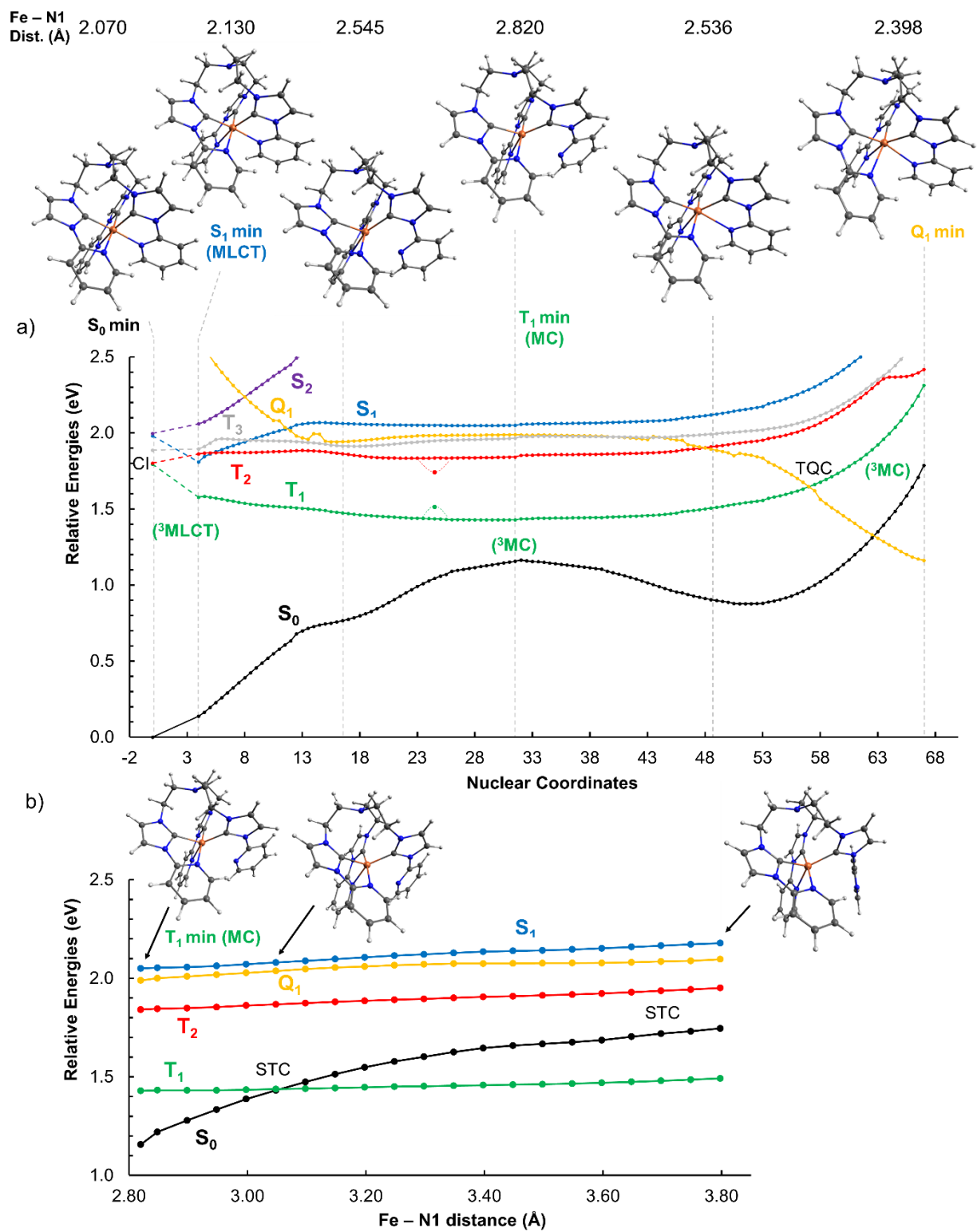

**Figure 4.** a) Photochemical landscape of **C2**. Arbitrary nuclear coordinate values are used to label the structures obtained along the decay path. The nuclear coordinates 0, 4.0, 31.5 and 67.0

correspond to the $S_0$ min, $S_1$ min ($^1$MLCT), $T_1$ min ($^3$MC) and $Q_1$ min, respectively. $T_1$ and $T_2$ energies at the $T_2$ min ($^3$MC) structure (nuclear coordinates = 24.5) are also shown. b) Energies of the most relevant states mapped upon Fe-N1 bond stretching by means of relaxed scan calculations on the $T_1$ surface starting from the $T_1$ min ($^3$MC) geometry of **C2**. All energies are relative to the $S_0$ minimum. STC = singlet-triplet crossing, TQC = triplet-quintet crossing.

Conversely, the $T_2$-$S_0$ STC area is achieved at larger Fe-N1 elongations. It shall be noted that the Fe-N1 enlargement has a small impact on the triplet PESs but greatly destabilizes the ground state, giving rise to relatively planar triplet surfaces and unveiling the absence of significant energy barriers throughout the decay process.

Interestingly, the $T_1$ ($^3$MC) min geometry lies at a relative energy of 1.43 eV, similar to the 1.35 eV determined for *fac*-**C1** and higher in energy with respect to *mer*-**C1** (1.13 eV).[23] On the other hand, the $T_2$ ($^3$MC) state has a value of 1.74 eV at its corresponding optimized geometry (nuclear coordinates = 24.5), revealing little relaxation energy for the $T_2$ state (~0.10-0.15 eV) and thus validating the current description of the triplet PESs used to study the decay channels of the complexes under study.

It can also be readily seen that the $T_1$-$Q_1$ triplet-quintet crossing (TQC) is less accessible from the $T_1$ ($^3$MC) min structure as compared to the $S_0$-$T_1$ STC, since an energy barrier has to be surmounted in order to populate the quintet state, whereas the access to the $S_0$-$T_1$ STC is barrierless. Therefore, it becomes apparent that the PES landscape of **C2** can be considered globally equivalent to the one of *fac*-**C1** and is significantly different from the one of the *mer* isomer.

**Femtosecond transient absorption spectroscopy**

In order to understand the photophysics of **C1** and **C2**, their solutions in ACN are studied using transient absorption spectroscopy (TAS), under excitation with ≈50 fs pulses at 400 nm. In Fig. 5A, some relevant spectra at different time delays are reported for the

*fac*-only **C2**. The negative signal is the ground state bleach (GSB), i.e this band mirrors the depletion of the population of the ground state induced by the pump.

Such signal is broad, from 300 up to 475 nm, but two main peaks are detected at 427 nm and another, less intense and narrow, at 364 nm. ~~The first one, at 0.3 ps overlaps with the inverse Steady State Absorption SSA, (blue dash- dot line) in the region 410-450 nm, while outside of this interval, a broad positive signal, due to excited state absorption (ESA), reduces the negative intensity.~~ The first one at 0.3 ps overlaps with the inverse Steady State Absorption (SSA) at 410-450 nm. Outside of this interval however, the negative intensity is reduced by a broad positive signal, due to excited state absorption (ESA). Indeed, the smaller GSB amplitude as compared to the SSA for $\lambda < 380$ nm (25% reduction at 0.3 ps) is due to a hidden ESA for these wavelengths. ESA appears as a dominant positive signal for $\lambda > 475$ nm and even 460 nm for longer delay times (> 5 ps). This band is broad, extending from 475 to beyond 650 nm at 0.3 ps, narrows down and blue-shifts for longer delay times.

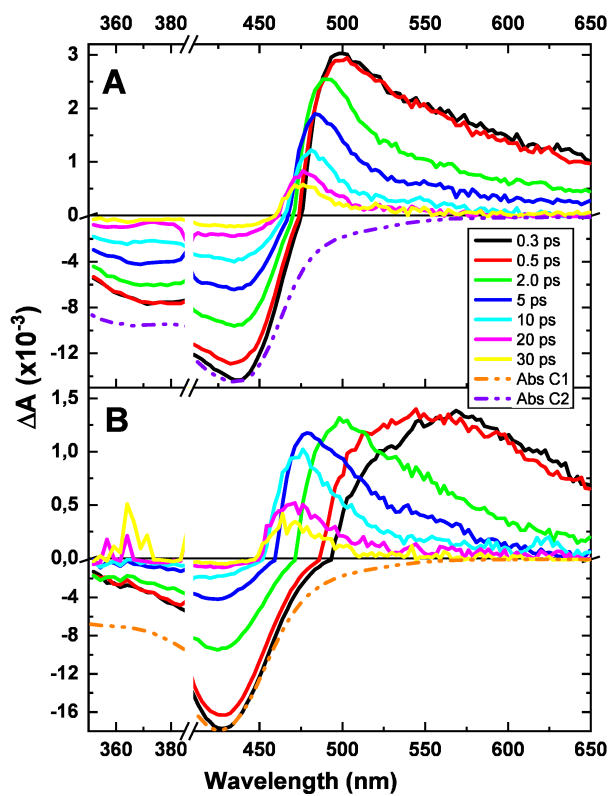

**Figure 5.** Transient absorption spectra at relevant time of (A) *fac*-only **C2** and (B) **C1** *fac/mer* 1/14 in $CH_3CN$ after excitation at 400 nm. Data at 390-415 nm are not shown (pump laser scattering). Dash-dotted lines are the respective inverted and scaled ground state absorption spectra. Note the different scales for positive and negative signals.

This is highlighted when the ΔA spectra are normalized at the ESA peak (Fig. S21). Spectral evolution stops only after 20 ps, when a ≈50 nm narrow spectrum with a maximum at 475 nm remains. Note that these ΔA spectra are significantly different from the ones we reported for iron complexes with tridentate NHC-based ligands.[10,11,12] Here, the extinction coefficients of the ESA transitions are 4, or even ≈10 (cf. **C1** below), times smaller than for the $S_0 \rightarrow {}^1MLCT$ transitions dominating the GSB, while for the iron complexes with tridentate ligands, the ratio is not larger than 3. In addition, complexes with tridentate ligands display additional positive ESA bands below 450 nm, which are not observed, or probably hidden in the GSB signal for **C1** and **C2**, here.

The result of the continuous spectral shifts and narrowing is a strong wavelength dependence of the kinetic traces in the ESA region (Fig. 5A, and S20, S21).

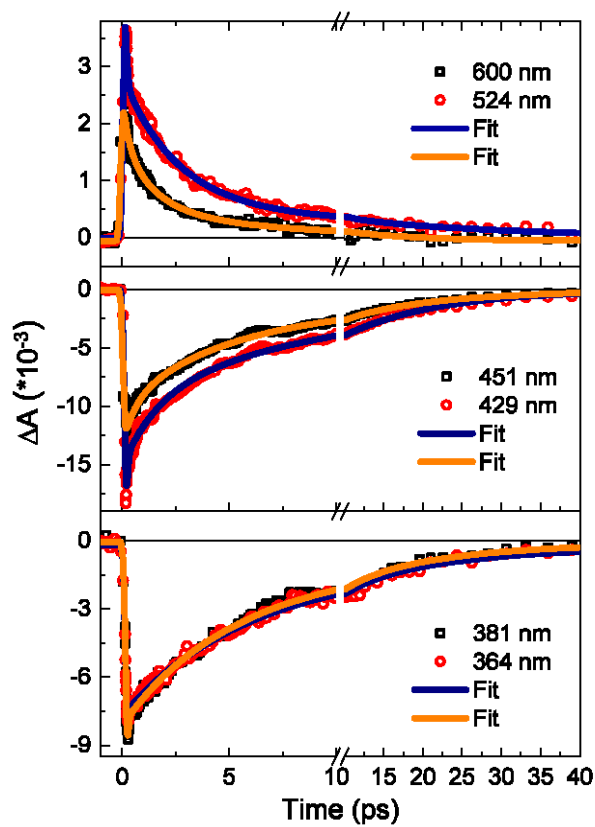

**Figure 6.** Kinetic traces (dots) and relative fits (line) of **C2** in different wavelength regions: ESA in the top panel, main GSB in the centre and short wavelength GSB in the bottom.

Kinetic traces at >550 nm decay faster than for shorter wavelengths, due to the presence of a sub-picosecond component. A detailed inspection of the kinetic traces on a semi-log scale (Fig. S20) allows to identify three characteristic wavelength-dependent lifetimes describing all the kinetic traces and spectral evolution. They are clearly distinct, and the three slopes of the semi-log traces allow to estimate their values: $\tau_1$ = 0.2-0.5 ps, $\tau_2$= 3-4 ps, $\tau_3$= 15-20 ps. These are the used for and refined by single wavelength fits as explained in SI. The agreement between the fitted curves and the experimental data is very good as highlighted in Fig. 6, and remaining residuals are due to the noise level (pump intensity fluctuations).

While $\tau_1$ is attributed to electronic relaxation from the optically excited $^1$MLCT state and subsequent vibrational relaxation, the occurrence of two distinct excited state lifetimes is unusual for NHC-Fe(II) complexes. Indeed, previous reports agreed in the description of a sequential relaxation process attributing one single ps lifetime $\tau_2$ to the population of the $^3$MLCT state (Fig. 7A):

$$^1MLCT^+ \xrightarrow{<IRF} \begin{cases} ^1MLCT \\ ^3MLCT^+ \end{cases} \xrightarrow{\tau_1} {}^3MCLT \xrightarrow{\tau_2} S_0 .$$

In the present case, for **C2** containing only the *fac* isomer, the question arises, whether $\tau_2$ reflects an internal conversion between two excited state X and Y, the latter decaying in a second stage with $\tau_3$ (Fig. 7B), or whether states X and Y decay in parallel with these two lifetimes (FigC).

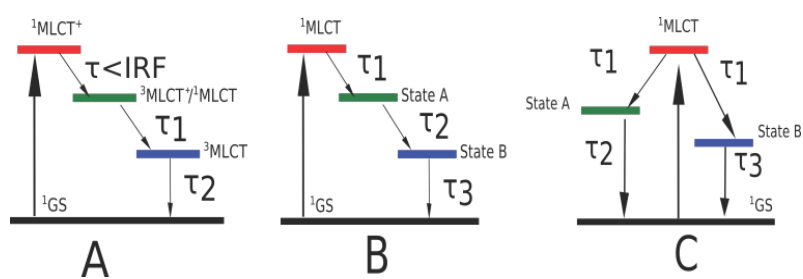

**Figure 7.** Excited state relaxation schemes in the presence of a single dominant ps excited state (A), or for the co-existence of 2 states labelled X and Y (schemes B and C), as found for ~~bidentate-ligated~~ **C1** and **C2** here.

A partial answer comes from an inspection of the time-resolved $\Delta$A spectra. Indeed, $\tau_3$ is associated with the long-lived quasi-static narrow ESA component dominating the spectra for t$\geq$ 10 ps, characterising state B (Fig. 5). In Fig. S19B, we display the results of a subtraction approach that removes the smoothed and normalized $\Delta$A*(t=15.0 ps) spectrum from the normalized $\Delta$A* at earlier times (the '*' denotes normalisation to the max. ESA signal). To a good degree of approximation, one may consider these double difference spectra $\Delta(\Delta A*(t)) = \Delta A*(t) - \Delta A*(t=15.0\ ps)$, as the time-dependent spectra of the earlier time populations, associated with $\tau_1$ ($^1$MLCT) and $\tau_2$ (state X). For delays t> 0.5 ps, the ESA part of $\Delta(\Delta A*(t))$, representative of state X, is much broader than the ESA of state Y. It displays only a slight dynamic blue-shift, so that a quasi-isobestic point appears at 485 nm, for t> 1.6 ps, with the GSB part of $\Delta(\Delta A*(t))$.

In Fig. 8, we summarize the results of the 3-exp. fits (see SI for details) performed for 20 single wavelengths.

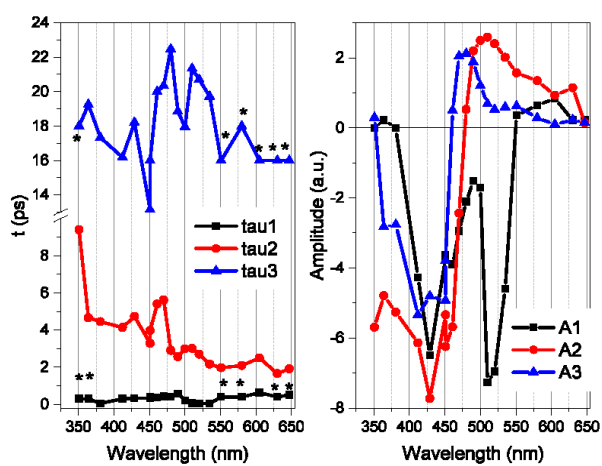

**Figure 8.** Results of single wavelength 3-exp. fits of **C2**. Left: Decay times vs wavelength. Right: Decay time associated amplitudes vs wavelength. Values are from table S3. '*' denote lifetime values kept fixed for fit convergence.

As anticipated above, at longer wavelengths (>550 nm) the ESA band is dominated by the amplitudes of $\tau_1$ ($^1$MLCT) and $\tau_2$ (state X), $A_1$ and $A_2$, while $A_3$ (state Y) is negligible. $\tau_1$ has an average value of 0.4±0.1 ps. It appears both in the ESA and GSB decay and is related to intersystem crossing from the $^1$MLCT state and subsequent intra-molecular vibrational cooling in states X and Y. As such it appears as a rise time for the ESA in the 500-540 nm range ($A_1$<0). In the bleach region ($\lambda$<450 nm), $\tau_2$ and $\tau_3$ are only slightly wavelength dependent with average values of 4±1 ps and 18±3 ps, respectively, while $\tau_2$ effectively decreases down to ≈2.0 ps for the longest ESA wavelengths. In Table S4 all the parameters are reported, together with details for the fit procedure, used for meaningful convergence of the fits. The $A_3$ amplitude associated with $\tau_3$, agrees with the 15 ps spectrum of Fig. S21B. In the range 440-600 nm, it has a positive amplitude, while below it is always negative (GSB). This decay associated amplitude spectrum demonstrates that the decay of state Y leads to filling of the ground state (GS), in agreement with schemes A and B of Fig. $A_2$, the amplitude of $\tau_2$ agrees with the evolution described in Fig. S21: The spectral shape of $A_2$ is consistent with the long-wavelength ESA component revealed after subtraction of the 15 ps spectrum (Fig. S19B). Most importantly, $A_2$ has a significant amplitude (30-40% larger than $A_3$) throughout the entire GSB region.

The spectral shape of $A_2$ allows us to discriminate between the parallel and sequential excited state relaxation schemes (Fig. 7, B and C). In the sequential scheme, the population moves from state X to state Y without any ground state bleach recovery. The 4 ps component should thus appear as a rise time (neg. $A_2$) for the ESA of Y in the 450-500 nm range. This is partly true since $A_2$ is negative for λ< 470 nm, but not for the wavelengths of maximum ESA of state Y (470-490nm). Hence, a clear indication of state X giving rise to population of state Y is lacking. More importantly is the observation of the GSB recovery occurring dominantly with $\tau_2$ for λ< 470 nm. This is in contradiction to the sequential relaxation scheme, but fully in agreement with the parallel one. Indeed, both the decays of the population of X and Y lead to ground state recovery. ~~As far as C1 is concerned, even though this complex contains only 6.7% of the *fac* isomer, the transient absorption data show, at first sight, a high resemblance with those of C2 (Fig. 5B).~~ In spite of containing only 6.7% of the *fac* isomer, the transient absorption data for **C1** show a high resemblance to those of **C2** (Fig. 5B). The main differences concern a smaller ratio of the ESA to GSB amplitude (1:10), and in a more pronounced evolution of the ESA in the early 2-5 ps. The ESA and GSB kinetics are significantly faster than for **C2**, as can be seen from Fig. 9, and from Figs. S24 and S26. Like **C2**, a detailed visual inspection of the kinetic traces in semi-log scales reveals three distinct lifetimes (Fig. S26): $\tau_1$ = 0.2-0.5 ps, $\tau_2$= 2-3 ps, $\tau_3$= 15-20 ps. Note that $\tau_2$ is shorter for **C1** than for **C2**.

A comparison of kinetic traces obtained for the 1:14 and a 1:8 mixture of *fac/mer* shows that they are identical within the experimental sensitivity (Fig. S27 and related discussion). Due to the limited dynamic range of TAS, and the similarity of the lifetimes of C1 and C2, the 10% minority contribution of *fac* isomers goes unnoticed in the data. We will therefore attribute the following excited state dynamics to *mer* isomers only.

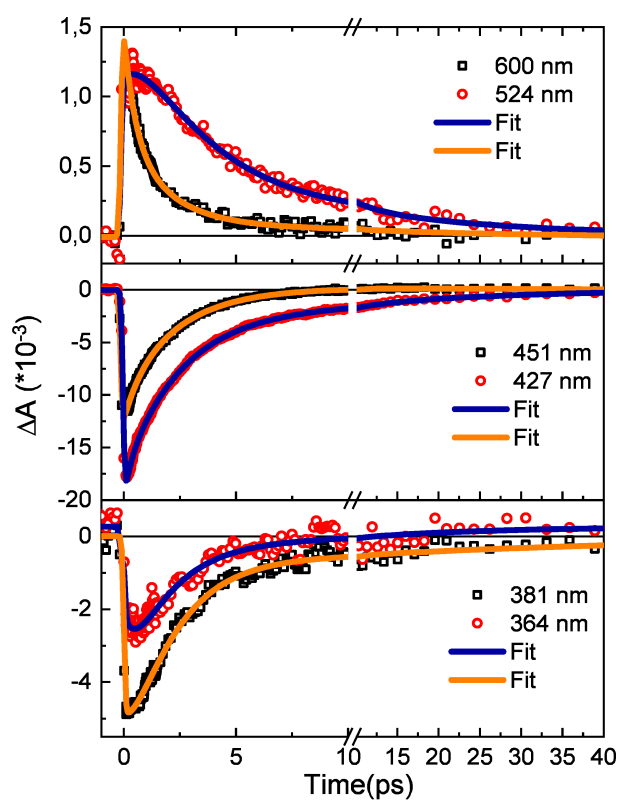

**Figure 9.** Kinetic traces (dots) and relative fits (line) of **C1** in different wavelength regions: ESA in the top panel, main GSB in the centre and short wavelength GSB in the bottom.

A similar analysis of the normalised time-resolved spectra, as performed for **C2**, allows to identify two distinct ESA bands and thus excited state populations, corresponding to lifetimes $\tau_2$ and $\tau_3$ (Fig. S25). As suggested by Fig.5B, the short wavelength ESA associated with $\tau_2$ is blue-shifting and decaying on that same timescale, leading to strongly wavelength dependent ESA kinetics and values for $\tau_2$, as can be readily seen from visual inspection (Fig. 9 top, Fig. S26 bottom). Global fitting is therefore not an option. Figs. S28 and S29 present the results of the 3-exp. fits carried out for individual wavelengths, and a detailed discussion is in the SI. Most importantly, the $\tau_2$ component associated with the short-lived long-wavelength ESA, characterising state X in Fig. 7, is the dominating channel for ground state recovery (Fig. S29). Also, the decay of state Y (lifetime $\tau_3$ and amplitude $A_3$), leads directly back into the ground state, since the latter is observed in the GSB recovery times, very clearly for (410-450 nm), where the signal-to-noise ratio is sufficiently large. We therefore conclude, that, as for **C2**, the parallel relaxation scheme of two distinct excited state populations (Fig. 7C) also applies for **C1.** If one wants to characterize the excited state lifetime by an average of the two populations, the experiments demonstrate that *mer* isomers have, on average, a 3-4 times shorter excited state lifetime than *fac*. It turns out that the faster overall decay of ESA and GSB of **C1** with respect to **C2** is mainly due to a larger relative amplitude for $\tau_2$, and its shorter value (2.0-2.5 ps). ~~Fig. S23 shows that, in the GSB region, the ratio of $A_2/A_3$ is ≈7:1, while for **C2**, this ratio is ≈4:3, on average~~. Fig. S23 shows on average a ratio of 7:1 ($A_2/A_3$) for the GSB region in comparison to 4:3 for **C2**. Other differences are in the smaller ESA amplitude and slightly red-shifted ESA spectra for *mer* with respect to *fac*.

**Overall interpretation of the excited-state decay**

With the computational results of the excited state manifold in mind, the first obvious question concerns the assignment of states X and Y and their location in the excited state landscape (Fig. 4). We can unambiguously rule out that the metal-centered quintet $^5T_2$ participates in the relaxation scheme since the latter is characterised by a sharp (10-15 nm broad) excited state absorption in the 340-360 nm range for tridentate ligands.[10,12] No sign of such an ESA is indicated by the ΔA spectra (Fig. 5). Furthermore, the calculated PES for both isomers of **C1** and **C2** consistently rule out the possible population of the quintet state due to its high energy. As a consequence, and in extension to previous reports,[10,34] we conclude that "3" is the minimum number of carbene bonds necessary to avoid the quintet population and hence photo-induced spin crossover in NHC-Fe(II) complexes.

One initial motivation for the design of bidentate ligands was the prospect to produce NHC complexes ~~with bite angles closer to 180°,~~ ensuring an ideally octahedral coordination of Fe, and thus increasing the ligand field, with respect to tridentate ligands. So far, to the best of our knowledge, no heteroleptic complex with tridentate ligands involving 3 carbene bonds has ~~not~~ been reported, making it difficult to identify a direct effect of the coordination geometry only. ~~The present study definitely points to an important modification of the excited state energy landscape since, for both *mer* and *fac* isomers, we observe the co-existence of two excited state populations, while in all homoleptic Fe(II)-NHC complexes with tridentate ligands complexes reported so far, a single excited state population was observed, always assigned to $^3$MLCT states.~~ Nevertheless, Dixon[36] and coworkers have investigated the effect of cyclometallation on the $^3$MLCT/$^3$MC relative energy levels in iron (II) coordinated with several tridentate phenylbipyridine (CNN) ligands. From this report, it appeared that the presence of a large portion of N-Fe-N coordination inducing a weaker ligand field effect in the complex dramatically stabilized the $^3$MC state while a notable destabilization was at play when N-Fe-C arrangements dominated. By analogy, the **C2** (fac) complex displays N-Fe-carbene arrangements exclusively explaining its longer excited state lifetime compared with **C1** (mer).

The present study certainly points to a considerable modification of the excited state energy landscape. Two excited state populations for both mer and fac isomers were observed, while for all reported homoleptic Fe(II)-NHC complexes with tridentate ligands, only a single excited state population, attributed to a $^3$MLCT, was found.[9,10,11,12,18,37]

The assignment of states X and Y in terms of their electronic character is difficult on the sole basis of the TAS data. We can, however, safely exclude them to be singlet states since these would display stimulated emission, i.e. negative ΔA on the low-energy side of the GSB bands. Triplet states, as the $^3$MLCT's reported for complexes with tridentate ligands, do not show sizeable stimulated emission since the radiative rate of the $^3$MLCT-$S_0$ transition has a ≈3 orders of magnitude smaller radiative rate than the allowed singlet-singlet transitions.[38] It is therefore most likely that the observed states X and Y are located on the triplet excited state PES presented in the computational studies. TAS could potentially differentiate the MC or MLCT character on the basis of the ESA spectra, however a clear fingerprint differentiating one from the other has only been reported for the very specific case of [Fe(bpy)(CN)$_4$]$^{2-}$, where the $^3$MLCT/MC crossover goes along with a strong solvatochromism.[35]

A possible interpretation of the time-resolved spectroscopy results and in particular the existence of a parallel decay model with a branching ratio between the fast and slow component that is strongly dependent on the isomerism, can be formulated taking into account the results from molecular modelling reported here and in previous contributions.[23]

The slower decay from the first triplet state may be explained by a closer examination of the PES. Figure 10 displays a schematic representation of the two-state decay model. Indeed, in the case of $T_1$ the system will be trapped in the $T_1$ minimum region where the $S_0$ PES is higher in energy than $T_1$ displaying a spin-crossover region. Hence, some vibrational cycles will be necessary to allow the crossing between the $S_0$ and $T_1$ states and subsequently transfer the population back to $S_0$. This process is thus ascribed to the slower decay component, $\tau_3$. On the other hand, a much larger region of the $T_2$ PES lies at higher energies with respect to $S_0$, having a much smaller spin-crossover area and thus being able to funnel the population to the available low-lying $S_0$ state in a much more efficient way, leading thus to shorter triplet excited-state lifetimes ($\tau_2$ component).

The difference in the branching ratio between the two paths in the case of *fac* and *mer* isomers can also be rationalized in terms of the different shape of the PES (Fig.10). In the case of *mer* (**C1**) one can observe an initially steep PES; furthermore, $T_1$ is significantly lower in energy than $T_2$ at the Franck-Condon region. The situation is totally different for the *fac* (**C2**) isomer where the PESs are globally flatter and the two triplet states are degenerated at Franck-Condon area. As such, in the case of the mer isomer a large population of $T_2$ should be expected in an ultrafast time scale, driving the system rapidly towards the central regions of the PES.[23] On the other hand, in the case of fac, and in the $\tau_1$ time-scale, i.e. close to the Franck-Condon region, the population of the triplet manifold will distribute more equally between $T_1$ and $T_2$ states by means of internal conversions. This interpretation is based on the flatter PESs and the presence of accessible $T_1/T_2$ energy degeneracy areas predicted at the surroundings of the mentioned Franck-Condon region (see Figure 4), in clear contrast to the mer profiles.[23]

Taking into account that the present experimental and computational results clearly determine a minor relevance of the $T_2$-$T_1$ deactivation channel, the delay of the $T_2$-$S_0$ deactivation up to the few-picosecond scale (2-3 ps for **C1**, 4 ps for **C2**), may be ascribed to the small energy splitting between the $T_2$-$T_3$ states, which could lead to a $T_2$-$T_3$ non-adiabatic trap. One should consider that higher triplet states ($T_4$, $T_5$, and so on) are greatly destabilized by the Fe-N enlargement in both fac and mer isomers of **C1**.[23] The $T_3$-$S_0$ decay route has not been explored in the present work and therefore remains as a hypothesis for future studies.

The steeper PES calculated for the mer isomer is also in agreement with the shorter value observed for the $\tau_2$ component of this isomer (2-3 ps) as compared to the $\tau_2$ measured for fac (4 ps). This measurement strengthens the combined experimental/theoretical interpretation shown in Figure 10.

Theoretical determinations of the triplet ESA spectra support the present interpretation. Although the calculation of ESA spectra has more technical difficulties as compared to the steady-state ones, and hence the results should be taken only as semi-quantitative, the overall experimentally observed trend of separate ESA bands with different lifetimes is confirmed by our simulations. Figures S17 to S19 show the simulated ESA for the $T_1{\rightarrow}T_X$ and $T_2{\rightarrow}T_X$ excitations computed for both **C2** and *mer*-**C1** complexes along some important points of their respective PES. At all Fe-N1 distances, the lowest-energy $T_2{\rightarrow}T_X$ absorptions appear at longer wavelengths with respect to the lowest-energy $T_1{\rightarrow}T_X$ excitations, explaining the apparent blue shift at the 450-500 nm region observed in the transient absorption spectra (see Figure 5) because of the faster $T_2$ decay. In other words, the faster disappearance of the $T_2{\rightarrow}T_X$ absorptions causes the mentioned apparent blue shift displayed in Figure 5.

Importantly, and in contrast with what is usually taken as a model, it should be noted that the two low-lying triplet states cannot clearly be identified within the diabatic notation as MC or MLCT. Instead, their nature changes adiabatically along the PES, and following the iron nitrogen enlargement, leads to a progressive deactivation rather than a sharp non-adiabatic transition from a ligand- to a metal-centred surface.

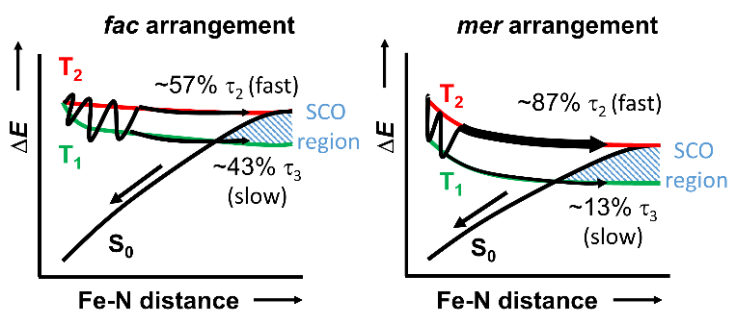

**Fig. 10.** Schematic interpretation of the $\tau_2$ and $\tau_3$ excited-state components on the basis of the singlet and triplet PES topologies of the *fac* and *mer* arrangements.

METHODOLOGY

All general experimental considerations may be found in the ESI.†

**Synthesis**

**Methylimidazoliumpyridine (L1)**[25]

2-pyridylimidazole (200 mg, 1.38 mmol) and iodomethane (1.76mg, 0.77 mL, 12.40 mmol) were heated one hour in 2 mL of acetonitrile at 100°C under microwaves irradiation (60W). After evaporation of solvent, the resulting mixture was dissolved in water. Then the metathesis was made by addition of saturated solution of KPF$_6$. Solid was filtered, washed twice with Et$_2$O and dried under vacuum to give yellow powder (67 %). The obtained NMR data for the ligand L1: $^1$H NMR (400 MHz, DMSO-d$_6$) : $\delta$ = 10.01, (s, 1H), 8.65 (ddd, $J$ = 4.8, 1.8, 0.8 Hz, 1H), 8.50 (t, $J$ = 2.0 Hz, 1H), 8.22 (td, $J$ =

7.6, 2.0 Hz, 1H), 7.99 (dt, $J$ = 13.6 Hz, 1H), 7.94(t, $J$ = 1.8 Hz, 1H), 7.65 (ddd, $J$ = 7.6, 5.0, 0.8 Hz, 1H), 3.97 (s, 3H) ppm.

**[Fe$^{II}$(MeIm-Py)](PF$_6$)$_2$ (C1)**

Using Schlenk techniques, the ligand **L1** (600 mg, 1.97 mmol) and FeCl$_2$, (82 mg, 0.65 mmol) were dissolved in 4 mL of DMF. Potassium tert-butoxide (243 m, 2.16 mmol) was then added. The dark red solution was stirred at rt for 10 min. After evaporation of DMF, the residue was purified by column chromatography using water/acetone/KNO$_3$ (6/3/1) as eluent. The red fractions were evaporated and a saturated solution of KPF$_6$ was added. The obtained solid was filtered, washed twice with Et$_2$O and dried under vacuum to give a non-separable mixture of both *fac*- and *mer*-**C1** in a 1:14 ratio (41 %) as an orange solid.

*mer*-**C1**: $^1$H NMR (400 MHz, CD$_3$CN) : $\delta$ = 8.22 (d, $J$ = 2.2 Hz, 1H), 8.15 (d, $J$ = 2.2 Hz, 1H), 8.11–8.08 (m,1H), 8.06 (d, $J$ = 2.3 Hz, 1H), 7.98 (d, $J$ = 8.3 Hz , 1H) 7.87–7.80 (m, 3H), 7.69 (d, $J$ = 8.2 Hz, 1H), 7.66 (d, $J$ = 8.3 Hz, 1H), 7.41 (d, $J$ = 2.2 Hz, 1H), 7.38 (d, $J$ = 2.2 Hz, 1H), 7.32 (dd, $J$ = 5.8, 0.6 Hz, 1H), 7.29 (td, $J$ = 5.6, 1.2 Hz, 1H), 7.19 (d, $J$ = 2.3 Hz, 1H), 6.91–6.89 (m, 2H), 6.85–6.83 (dd, $J$ = 5.8, 0.7 Hz, 1H), 3.25 (s, 3H), 2.82 (s, 3H), 2.70 (s, 3H) ppm. $^{13}$C NMR (100 MHz, CD$_3$CN) : $\delta$ = 209.5, 205.1, 204.5, 157.1, 156.3, 155.9, 155.7, 155.3, 152.5, 141.5, 140.2, 139.1, 129.0, 128.9, 128.8, 124.7, 123.4, 122.7, 118.9, 118.0, 117.8, 113.2, 112.4, 112.1, 37.0, 36.0, 35.8 ppm.

*fac*-**C1**: $^1$H NMR (400 MHz, CD$_3$CN) : $\delta$ = 8.11 (d, $J$ = 1.6 Hz, 1H), 8.11–8.06 (m,1H), 7.91 (d, $J$ = 8.3 Hz, 1H), 7.24–7.21 (m, 1H), 7.18 (d, $J$ = 2.5 Hz, 1H), 7.07 (dd, $J$ = 5.7, 0.7 Hz, 1H), 2.97 (s, 3H) ppm. $^{13}$C NMR (100 MHz, CD$_3$CN) : $\delta$ = 200.3, 151.1, 140.7, 128.5, 124.1, 118.2, 112.9, 36.7ppm. HR-MS calcd for C$_{27}$H$_{27}$FeN$_9$P$_2$F$_{12}$ m/z: 266.5864 [M - 2PF$_6$]$^{2+}$, found 266.5873.

**Tris(2-imidazoliumpyridineethyl)amine (L2)**

Tris(2-chloroethyl)amine (100 mg, 0.41 mmol) and 2-pyridylimidazole (362 mg, 2.46 mmol) were reacted for one hour in 2 mL of isopropanol at 160°C under microwave irradiation (70W). After evaporation of the solvent, the resulting mixture was dissolved in water. After addition of a saturated solution of KPF$_6$ the obtained solid was filtered, washed twice with Et$_2$O and dried under vacuum to give **L2** (98 % yield). $^1$H NMR (400 MHz, DMSO-d$_6$) : $\delta$ = 9.92 (s, 3H), 8.56 (dd, $J$ = 4.8, 1.0 Hz, 3H), 8.44 (t, $J$ = 2.0 Hz ,3H), 8.18 (td, $J$ = 7.6, 2.0 Hz ,6H), 7.92–7.89 (m, 6H), 7.62 (dd, $J$ = 7.6, 4.8 Hz, 3H), 4.40 (t, $J$ = 6.0 Hz, 6H), 3.17 (t, $J$ = 6.0 Hz, 6H) ppm. $^{13}$C NMR (100 MHz, DMSO-d$_6$): $\delta$ = 149.2, 146.1, 140.6, 135.0, 125.2, 132.7, 119.0, 113.9, 51.3, 46.4 ppm. HR-MS calcd for C$_{30}$H$_{33}$N$_{10}$P$_3$F$_{18}$ m/z: 823.2168 [M - PF$_6$]$^+$, found 823.2258.

**[Fe$^{II}$(N-Et-Im-Py)](PF$_6$)$_2$ (C2)**

Using Schlenk techniques, the ligand **L2** (400 mg, 0.42 mmol) and FeCl$_2$, (4 mg, 0.38 mmol) were dissolved in 15 mL of DMF. Potassium tert-butoxide (281 mg, 2.50 mmol) was then added. The dark red solution stirred at rt for 10 min. After evaporation of DMF, the resulting crude was purified by column chromatography using water/acetone/KNO$_3$ (6/3/1) as eluent. The red fractions were evaporated and a saturated solution of KPF$_6$ was added. The solid was filtered, washed twice with Et$_2$O and dried under vacuum to give **C2** as an orange solid (15 % yield). $^1$H NMR (400 MHz, CD$_3$CN) : $\delta$ = 8.16 (s, 3H), 8.06 (t, $J$ = 6.6 Hz, 6H), 7.90 (dd, $J$ = 8.6 Hz, 3H), 7.21–7.18 (m, 6H), 7.04 (d, $J$ = 5.6 Hz, 3H), 3.60 (d, $J$ = 14.6 Hz, 3H), 3.14 (d, $J$ = 13.6 Hz, 3H), 2.84 (t, $J$ = 14.6 Hz, 3H), 1.95 (t, 3H) ppm. $^{13}$C NMR (100 MHz, CD$_3$CN): $\delta$ = 207.5, 154.6, 150.7, 140.2, 127.2, 123.7, 119.2, 112.6, 63.8, 52.9 ppm. HR-MS calcd for C$_{30}$H$_{30}$FeN$_{10}$P$_2$F$_{12}$ m/z : 293.0997 [M - 2PF$_6$]$^{2+}$, found 293.1020.

**Computational methods**

The computational methods employed in this work have been fully detailed elsewhere[23] and only a brief description will be provided here. All computations have been performed with the GAUSSIAN 09 (D01 revision) software package.[39] The $S_0$ and $Q_1$ equilibrium geometries of **C2** have been obtained with the standard DFT/B3LYP method, whereas optimizations and MEPs of the $T_1$ state have been performed using the unrestricted DFT/HCTH407 method (hereafter, DFT/HCTH).[40] On the other hand, the $S_1$ and $T_2$ states have been optimized at the TD-DFT/HCTH level. All energies discussed in the present work have been obtained with the TD-DFT/HCTH method. The GAUSSIAN 09 standard optimization thresholds have been used in all minimizations except for the $S_0$ equilibrium structures of **C0**, *fac*-**C1**, *mer*-**C1** and **C2** shown in Fig.2 and Table S1. In these cases, very tight optimization thresholds (opt=verytight) have been used to provide highly accurate geometrical parameters. The energies of all electronic states reported and discussed in the present work have been obtained at the same level of theory, employing the TD-DFT/HCTH method on top of the optimized geometries using the methodologies described above. The absorption energies used to convolute the theoretical spectrum displayed in Figure S13 have been blue-shifted by 0.28 eV to facilitate the comparison with the experimental recordings.

The NTOs[41] have been obtained by post-processing the GAUSSIAN output using the Nancy_EX code.[42,43] The photochemical landscape displayed in Fig.4 has been built by combining MEP calculations (relaxing the $T_1$ state) starting from the $S_1$ ($^1$MLCT) min (nuclear coordinate = 4.0) and the $Q_1$ min geometry (nuclear coordinate = 67.0). Both MEP computations ended up at the $T_1$ ($^3$MC) min structure (nuclear coordinate = 31.5), identified in a separated calculation using the standard optimization algorithm. In addition, the T$_2$ ($^3$MC) min structure has a Fe-N distance of 2.725 Å and the T1 and T2 energies at this geometry are shown at the nuclear coordinate of 24.5. The Tamm-Dancoff approximation[44] has been used in all TD-DFT calculations and the 6-31+G(d,p) basis set has been used throughout. The solvent effects (acetonitrile) have been included in all optimization and single-point computations by means of the polarizable continuum model (PCM) method, employing the GAUSSIAN 09 default settings.

**Femtosecond transient absorption spectroscopy**

In the transient absorption setup, an amplified 5kHz Ti: sapphire laser generates 30 fs pulses to pump a commercial optical parametric amplifier (TOPAS: Light Conversion), from which 60 fs pulses at 400 nm are derived to excite the samples. The white light continuum probe beam is generated in a 2 mm thick CaF$_2$ crystal and split in two: the probe that is sent through the sample, and a reference beam is used for measuring and compensating the white-light intensity fluctuations. The polarization of the probe beam is set at magic angle (54.7°) with respect to the pump.

A 1 mm path length cuvette in fused silica contains the complexes dissolved in CH$_3$CN. Time-resolved spectra are acquired in the range 350-650 nm as a function of pump-probe delay, by a Peltier-cooled CCD with 220Hz acquisition rate. A solvent-only sample is measured, and the data are processed to (i) remove the background at negative delay times, (ii) the solvent Raman signal and the coherent interactions of the pump and probe in the cell, (iii) to correct for the group velocity dispersion of the probe beam, characterized in the solvent-only data set. The temporal resolution is characterized by the 70±5 fs FWHM of the solvent Raman response. For more details see Ref.[45]. The multi-exponential fit procedure is described in the SI.

**CONCLUSIONS**

In the present contribution, we report the synthesis and the detailed experimental and computational characterization of an original iron-carbene bidendate compound. The complex is obtained with a 100% *fac* isomerism, in contrast to previous conventional synthetic procedures which only lead to mixtures dominated by the *mer* isomer. The excited-state lifetimes of the pure *fac* compound have been compared with those of the *mer* isomer. It is shown that, as it was the case for complexes with tridentate ligands reported before, the inclusion of carbene moieties increases the MLCT lifetime from the subpicosecond range to a maximum of about 20 ps.

By producing the constrained *fac*-only complex with bidentate ligands, we have confirmed large differences in the photophysics induced by bidentate and tridentate. Our transient absorption spectroscopy measurements and the computational results unveil crucial differences in the photophysics of the *fac* and *mer* isomers. We show that complexes with bidentate ligands decay to the ground state via a parallel two-state mechanism composed by a fast and a slow component. The former component is ascribed to the population of the $T_2$ state whereas the latter is ascribed to the $T_1$ state. Differences in the excited-state lifetime are interpreted based on a large spin-crossover region in $T_1$ as compared to $T_2$, which will act as a more efficient trap for the triplet state slowing down its decay.

While in the case of the *fac* isomer the two pathways appear as equally probable, a preference for the fastest decay is observed in the *mer* dominated mixture, hence globally leading to a faster decay for the latter.

Our results clearly confirm the peculiarity of the photophysics of iron complexes with bidentate ligands, especially the crucial role of *fac*/*mer* isomerism in driving the overall decay process, illustrating how the simple diabatic picture relying in the competition between well separated MC and MLCT states is largely insufficient to provide a full coherent explanation of the time resolved photophysics.

ASSOCIATED CONTENT

NMR, spectroscopic and electrochemical data for complexes as well as additional computational and photophysical details can be found in the Supporting Information.

ACKNOWLEDGMENT

The project is funded by the French Agence Nationale de la Recherche (ANR-16-CE07-0013-02). The Nancy team is grateful to S. Parant for UV spectroscopy and F. Lachaud for mass spectrometry. A.F.-M. acknowledges the Région Grand Est government (France) for the financial support. The Strasbourg group acknowledges financial support through the Labex NIE.

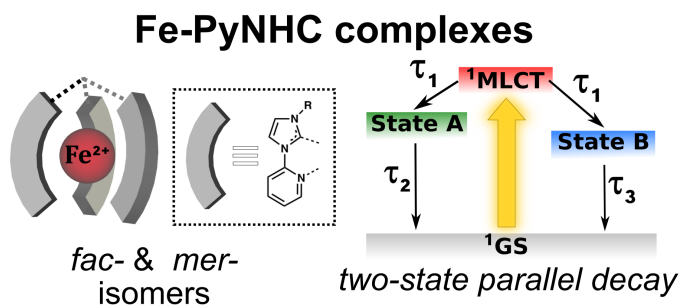

**Fe-PyNHC complexes**

fac- & mer- isomers

two-state parallel decay

From combined experimental and computational investigations, a mechanism based on the differential trapping of the triplet states in spin-crossover regions is proposed for the first time to explain the impact of the fac/mer isomerism on the overall excited-state dynamics of iron (II) complexes with bidentate Pyridyl-NHC ligands.